\documentclass[12pt]{article}
\input epsf

\makeatletter
\newcount\@hour\newcount\@minute\chardef\@x10\chardef\@xv60
\def\tcitime{
\def\@time{%
  \@minute\time\@hour\@minute\divide\@hour\@xv
  \ifnum\@hour<\@x 0\fi\the\@hour:%
  \multiply\@hour\@xv\advance\@minute-\@hour
  \ifnum\@minute<\@x 0\fi\the\@minute
  }}%
\def\QCTOpt[#1]#2{%
  \def\QCTOptB{#1}
  \def\QCTOptA{#2}
}
\def\QCTNOpt#1{%
  \def\QCTOptA{#1}
  \let\QCTOptB\empty
}
\def\Qct{%
  \@ifnextchar[{%
    \QCTOpt}{\QCTNOpt}
}
\def\QCBOpt[#1]#2{%
  \def\QCBOptB{#1}
  \def\QCBOptA{#2}
}
\def\QCBNOpt#1{%
  \def\QCBOptA{#1}
  \let\QCBOptB\empty
}
\def\Qcb{%
  \@ifnextchar[{%
    \QCBOpt}{\QCBNOpt}
}
\def\PrepCapArgs{%
  \ifx\QCBOptA\empty
    \ifx\QCTOptA\empty
      {}%
    \else
      \ifx\QCTOptB\empty
        {\QCTOptA}%
      \else
        [\QCTOptB]{\QCTOptA}%
      \fi
    \fi
  \else
    \ifx\QCBOptA\empty
      {}%
    \else
      \ifx\QCBOptB\empty
        {\QCBOptA}%
      \else
        [\QCBOptB]{\QCBOptA}%
      \fi
    \fi
  \fi
}
\newcount\GRAPHICSTYPE
\GRAPHICSTYPE=\z@
\def\GRAPHICSPS#1{%
 \ifcase\GRAPHICSTYPE%\GRAPHICSTYPE=0
   \special{ps: #1}%
 \or%\GRAPHICSTYPE=1
   \special{language "PS", include "#1"}%
 \fi
}%
\def\graffile#1#2#3#4{%
    \leavevmode
    \raise -#4 \BOXTHEFRAME{%
        \hbox to #2{\raise #3\hbox{\null #1}}}%
}%
\def\draftbox#1#2#3#4{%
 \leavevmode\raise -#4 \hbox{%
  \frame{\rlap{\protect\tiny #1}\hbox to #2%
   {\vrule height#3 width\z@ depth\z@\hfil}%
  }%
 }%
}%
\newcount\draft
\draft=\z@

\newif\ifwasdraft
\wasdraftfalse

\def\GRAPHIC#1#2#3#4#5{%
 \ifnum\draft=\@ne\draftbox{#2}{#3}{#4}{#5}%
  \else\graffile{#1}{#3}{#4}{#5}%
  \fi
 }%
\def\addtoLaTeXparams#1{%
    \edef\LaTeXparams{\LaTeXparams #1}}%
\newif\ifBoxFrame \BoxFramefalse
\newif\ifOverFrame \OverFramefalse
\newif\ifUnderFrame \UnderFramefalse

\def\BOXTHEFRAME#1{%
   \hbox{%
      \ifBoxFrame
         \frame{#1}%
      \else
         {#1}%
      \fi
   }%
}

\def\doFRAMEparams#1{\BoxFramefalse\OverFramefalse\UnderFramefalse\readFRAMEparams#1\end}%
\def\readFRAMEparams#1{%
 \ifx#1\end%
  \let\next=\relax
  \else
  \ifx#1i\dispkind=\z@\fi
  \ifx#1d\dispkind=\@ne\fi
  \ifx#1f\dispkind=\tw@\fi
  \ifx#1t\addtoLaTeXparams{t}\fi
  \ifx#1b\addtoLaTeXparams{b}\fi
  \ifx#1p\addtoLaTeXparams{p}\fi
  \ifx#1h\addtoLaTeXparams{h}\fi
  \ifx#1X\BoxFrametrue\fi
  \ifx#1O\OverFrametrue\fi
  \ifx#1U\UnderFrametrue\fi
  \ifx#1w
    \ifnum\draft=1\wasdrafttrue\else\wasdraftfalse\fi
    \draft=\@ne
  \fi
  \let\next=\readFRAMEparams
  \fi
 \next
 }%
\def\IFRAME#1#2#3#4#5#6{%
      \bgroup
      \let\QCTOptA\empty
      \let\QCTOptB\empty
      \let\QCBOptA\empty
      \let\QCBOptB\empty
      #6%
      \parindent=0pt%
      \leftskip=0pt
      \rightskip=0pt
      \setbox0 = \hbox{\QCBOptA}%
      \@tempdima = #1\relax
      \ifOverFrame
          \typeout{This is not implemented yet}%
          \show\HELP
      \else
         \ifdim\wd0>\@tempdima
            \advance\@tempdima by \@tempdima
            \ifdim\wd0 >\@tempdima
               \textwidth=\@tempdima
               \setbox1 =\vbox{%
                  \noindent\hbox to \@tempdima{\hfill\GRAPHIC{#5}{#4}{#1}{#2}{#3}\hfill}\\%
                  \noindent\hbox to \@tempdima{\parbox[b]{\@tempdima}{\QCBOptA}}%
               }%
               \wd1=\@tempdima
            \else
               \textwidth=\wd0
               \setbox1 =\vbox{%
                 \noindent\hbox to \wd0{\hfill\GRAPHIC{#5}{#4}{#1}{#2}{#3}\hfill}\\%
                 \noindent\hbox{\QCBOptA}%
               }%
               \wd1=\wd0
            \fi
         \else
            \ifdim\wd0>0pt
              \hsize=\@tempdima
              \setbox1 =\vbox{%
                \unskip\GRAPHIC{#5}{#4}{#1}{#2}{0pt}%
                \break
                \unskip\hbox to \@tempdima{\hfill \QCBOptA\hfill}%
              }%
              \wd1=\@tempdima
           \else
              \hsize=\@tempdima
              \setbox1 =\vbox{%
                \unskip\GRAPHIC{#5}{#4}{#1}{#2}{0pt}%
              }%
              \wd1=\@tempdima
           \fi
         \fi
         \@tempdimb=\ht1
         \advance\@tempdimb by \dp1
         \advance\@tempdimb by -#2%
         \advance\@tempdimb by #3%
         \leavevmode
         \raise -\@tempdimb \hbox{\box1}%
      \fi
      \egroup%
}%
\def\DFRAME#1#2#3#4#5{%
 \begin{center}
     \let\QCTOptA\empty
     \let\QCTOptB\empty
     \let\QCBOptA\empty
     \let\QCBOptB\empty
     \ifOverFrame 
        #5\QCTOptA\par
     \fi
     \GRAPHIC{#4}{#3}{#1}{#2}{\z@}
     \ifUnderFrame 
        \par #5\QCBOptA
     \fi
 \end{center}%
 }%
\def\FFRAME#1#2#3#4#5#6#7{%
 \begin{figure}[#1]%
  \let\QCTOptA\empty
  \let\QCTOptB\empty
  \let\QCBOptA\empty
  \let\QCBOptB\empty
  \ifOverFrame
    #4
    \ifx\QCTOptA\empty
    \else
      \ifx\QCTOptB\empty
        \caption{\QCTOptA}%
      \else
        \caption[\QCTOptB]{\QCTOptA}%
      \fi
    \fi
    \ifUnderFrame\else
      \label{#5}%
    \fi
  \else
    \UnderFrametrue%
  \fi
  \begin{center}\GRAPHIC{#7}{#6}{#2}{#3}{\z@}\end{center}%
  \ifUnderFrame
    #4
    \ifx\QCBOptA\empty
      \caption{}%
    \else
      \ifx\QCBOptB\empty
        \caption{\QCBOptA}%
      \else
        \caption[\QCBOptB]{\QCBOptA}%
      \fi
    \fi
    \label{#5}%
  \fi
  \end{figure}%
 }%
\newcount\dispkind%
\def\FRAME#1#2#3#4#5#6#7#8{%
 \ifnum\draft=\@ne
   \wasdrafttrue
 \else
   \wasdraftfalse%
 \fi
 \def\LaTeXparams{}%
 \dispkind=\z@
 \def\LaTeXparams{}%
 \doFRAMEparams{#1}%
 \ifnum\dispkind=\z@\IFRAME{#2}{#3}{#4}{#7}{#8}{#5}\else
  \ifnum\dispkind=\@ne\DFRAME{#2}{#3}{#7}{#8}{#5}\else
   \ifnum\dispkind=\tw@
    \edef\@tempa{\noexpand\FFRAME{\LaTeXparams}}%
    \@tempa{#2}{#3}{#5}{#6}{#7}{#8}%
    \fi
   \fi
  \fi
  \ifwasdraft\draft=1\else\draft=0\fi{}%
 }%
\def\TEXUX#1{"texux"}

\long\def\QQQ#1#2{%
     \long\expandafter\def\csname#1\endcsname{#2}}%
\@ifundefined{QTP}{\def\QTP#1{}}{}
\@ifundefined{Qlb}{}{}
\@ifundefined{Qlt}{}{}
\long\def\QQA#1#2{}%
\def\QTR#1#2{{\csname#1\endcsname #2}}%(gp) Is this the best?
\long\def\TeXButton#1#2{#2}%
\def\EXPAND#1[#2]#3{}%
\def\NOEXPAND#1[#2]#3{}%
\def\LaTeXparent#1{}%
\def\ChildStyles#1{}%
\def\ChildDefaults#1{}%
\def\QTagDef#1#2#3{}%
\def\QQfnmark#1{\footnotemark}

\@ifundefined{INDEX}{\def\INDEX#1#2{}{}}{}%
\@ifundefined{SUBINDEX}{\def\SUBINDEX#1#2#3{}{}{}}{}%
\def\initial#1{\bigbreak{\raggedright\large\bf #1}\kern 2\p@
   \penalty3000}%
\@ifundefined{ZZZ}{}{\makeatletter\input gnuindex.sty\makeatother\makeindex\makeatletter}%
\@ifundefined{abstract}{%
 \def\abstract{%
  \if@twocolumn
   \section*{Abstract (Not appropriate in this style!)}%
   \else \small 
   \begin{center}{\bf Abstract\vspace{-.5em}\vspace{\z@}}\end{center}%
   \quotation 
   \fi
  }%
 }{%
 }%
\@ifundefined{endabstract}{\def\endabstract
  {\if@twocolumn\else\endquotation\fi}}{}%
\@ifundefined{maketitle}{\def\maketitle#1{}}{}%
\@ifundefined{affiliation}{\def\affiliation#1{}}{}%
\@ifundefined{proof}{}{}%
\@ifundefined{endproof}{}{}%
\@ifundefined{newfield}{\def\newfield#1#2{}}{}%
\@ifundefined{chapter}{\def\chapter#1{\par(Chapter head:)#1\par }%
 \newcount\c@chapter}{}%
\@ifundefined{part}{\def\part#1{\par(Part head:)#1\par }}{}%
\@ifundefined{section}{\def\section#1{\par(Section head:)#1\par }}{}%
\@ifundefined{subsection}{\def\subsection#1%
 {\par(Subsection head:)#1\par }}{}%
\@ifundefined{subsubsection}{\def\subsubsection#1%
 {\par(Subsubsection head:)#1\par }}{}%
\@ifundefined{paragraph}{\def\paragraph#1%
 {\par(Subsubsubsection head:)#1\par }}{}%
\@ifundefined{subparagraph}{\def\subparagraph#1%
 {\par(Subsubsubsubsection head:)#1\par }}{}%
\@ifundefined{therefore}{}{}%
\@ifundefined{backepsilon}{}{}%
\@ifundefined{yen}{}{}%
\@ifundefined{registered}{%
   \def\registered{\relax\ifmmode{}\r@gistered
                    \else$\m@th\r@gistered$\fi}%
 \def\r@gistered{^{\ooalign
  {\hfil\raise.07ex\hbox{$\scriptstyle\rm\text{R}$}\hfil\crcr
  \mathhexbox20D}}}}{}%
\@ifundefined{Eth}{}{}%
\@ifundefined{eth}{}{}%
\@ifundefined{Thorn}{}{}%
\@ifundefined{thorn}{}{}%
\@ifundefined{degree}{}{}%
\newdimen\theight
\def\Column{%
 \vadjust{\setbox\z@=\hbox{\scriptsize\quad\quad tcol}%
  \theight=\ht\z@\advance\theight by \dp\z@\advance\theight by \lineskip
  \kern -\theight \vbox to \theight{%
   \rightline{\rlap{\box\z@}}%
   \vss
   }%
  }%
 }%
\def\qed{%
 \ifhmode\unskip\nobreak\fi\ifmmode\ifinner\else\hskip5\p@\fi\fi
 \hbox{\hskip5\p@\vrule width4\p@ height6\p@ depth1.5\p@\hskip\p@}%
 }%
\def\miss{\hbox{\vrule height2\p@ width 2\p@ depth\z@}}%
\def\tcol#1{{\baselineskip=6\p@ \vcenter{#1}} \Column}  %
\def\newfmtname{LaTeX2e}
\def\chkcompat{%
   \if@compatibility
   \else
     \usepackage{latexsym}
   \fi
}

\ifx\fmtname\newfmtname
  \DeclareOldFontCommand{\rm}{\normalfont\rmfamily}{\mathrm}
  \DeclareOldFontCommand{\sf}{\normalfont\sffamily}{\mathsf}
  \DeclareOldFontCommand{\tt}{\normalfont\ttfamily}{\mathtt}
  \DeclareOldFontCommand{\bf}{\normalfont\bfseries}{\mathbf}
  \DeclareOldFontCommand{\it}{\normalfont\itshape}{\mathit}
  \DeclareOldFontCommand{\sl}{\normalfont\slshape}{\@nomath\sl}
  \DeclareOldFontCommand{\sc}{\normalfont\scshape}{\@nomath\sc}
  \chkcompat
\fi

\def\alpha{\Greekmath 010B }%
\def\beta{\Greekmath 010C }%
\def\gamma{\Greekmath 010D }%
\def\delta{\Greekmath 010E }%
\def\epsilon{\Greekmath 010F }%
\def\zeta{\Greekmath 0110 }%
\def\eta{\Greekmath 0111 }%
\def\theta{\Greekmath 0112 }%
\def\iota{\Greekmath 0113 }%
\def\kappa{\Greekmath 0114 }%
\def\lambda{\Greekmath 0115 }%
\def\mu{\Greekmath 0116 }%
\def\nu{\Greekmath 0117 }%
\def\xi{\Greekmath 0118 }%
\def\pi{\Greekmath 0119 }%
\def\rho{\Greekmath 011A }%
\def\sigma{\Greekmath 011B }%
\def\tau{\Greekmath 011C }%
\def\upsilon{\Greekmath 011D }%
\def\phi{\Greekmath 011E }%
\def\chi{\Greekmath 011F }%
\def\psi{\Greekmath 0120 }%
\def\omega{\Greekmath 0121 }%
\def\varepsilon{\Greekmath 0122 }%
\def\vartheta{\Greekmath 0123 }%
\def\varpi{\Greekmath 0124 }%
\def\varrho{\Greekmath 0125 }%
\def\varsigma{\Greekmath 0126 }%
\def\varphi{\Greekmath 0127 }%

\def\nabla{\Greekmath 0272 }

\def\Greekmath#1#2#3#4{%
    \if@compatibility
        \ifnum\mathgroup=\symbold
           \mathchoice{\mbox{\boldmath$\displaystyle\mathchar"#1#2#3#4$}}%
                      {\mbox{\boldmath$\textstyle\mathchar"#1#2#3#4$}}%
                      {\mbox{\boldmath$\scriptstyle\mathchar"#1#2#3#4$}}%
                      {\mbox{\boldmath$\scriptscriptstyle\mathchar"#1#2#3#4$}}%
        \else
           \mathchar"#1#2#3#4% 
        \fi 
    \else 
        \ifnum\mathgroup=5 % For 2e
           \mathchoice{\mbox{\boldmath$\displaystyle\mathchar"#1#2#3#4$}}%
                      {\mbox{\boldmath$\textstyle\mathchar"#1#2#3#4$}}%
                      {\mbox{\boldmath$\scriptstyle\mathchar"#1#2#3#4$}}%
                      {\mbox{\boldmath$\scriptscriptstyle\mathchar"#1#2#3#4$}}%
        \else
           \mathchar"#1#2#3#4% 
        \fi     	    
	  \fi}

\newif\ifGreekBold  \GreekBoldfalse
\let\SAVEPBF=\pbf
\def\pbf{\GreekBoldtrue\SAVEPBF}%

\@ifundefined{theorem}{}{}
\@ifundefined{lemma}{}{}
\@ifundefined{corollary}{}{}
\@ifundefined{conjecture}{}{}
\@ifundefined{proposition}{}{}
\@ifundefined{axiom}{}{}
\@ifundefined{remark}{}{}
\@ifundefined{example}{}{}
\@ifundefined{exercise}{}{}
\@ifundefined{definition}{}{}

\@ifundefined{mathletters}{%
  \newcounter{equationnumber}  
  \def\mathletters{%
     \addtocounter{equation}{1}
     \edef\@currentlabel{\theequation}%
     \setcounter{equationnumber}{\c@equation}
     \setcounter{equation}{0}%
     \edef\theequation{\@currentlabel\noexpand\alph{equation}}%
  }
  
}{}

\@ifundefined{BibTeX}{%
    \def\BibTeX{{\rm B\kern-.05em{\sc i\kern-.025em b}\kern-.08em
                 T\kern-.1667em\lower.7ex\hbox{E}\kern-.125emX}}}{}%
\@ifundefined{AmS}%
    {\def\AmS{{\protect\usefont{OMS}{cmsy}{m}{n}%
                A\kern-.1667em\lower.5ex\hbox{M}\kern-.125emS}}}{}%
\@ifundefined{AmSTeX}{}{}%
\ifx\ds@amstex\relax
\makeatother\endinput% 2.09 compatability
\else
   \@ifpackageloaded{amstex}%
      {\message{amstex already loaded}\makeatother\endinput}
      {}
\fi
\let\DOTSI\relax
\def\RIfM@{\relax\ifmmode}%
\def\FN@{\futurelet\next}%
\newcount\intno@
\def\iint{\DOTSI\intno@\tw@\FN@\ints@}%
\def\iiint{\DOTSI\intno@\thr@@\FN@\ints@}%
\def\iiiint{\DOTSI\intno@4 \FN@\ints@}%
\def\idotsint{\DOTSI\intno@\z@\FN@\ints@}%
\def\ints@{\findlimits@\ints@@}%
\newif\iflimtoken@
\newif\iflimits@
\def\findlimits@{\limtoken@true\ifx\next\limits\limits@true
 \else\ifx\next\nolimits\limits@false\else
 \limtoken@false\ifx\ilimits@\nolimits\limits@false\else
 \ifinner\limits@false\else\limits@true\fi\fi\fi\fi}%
\def\multint@{\int\ifnum\intno@=\z@\intdots@                          %1
 \else\intkern@\fi                                                    %2
 \ifnum\intno@>\tw@\int\intkern@\fi                                   %3
 \ifnum\intno@>\thr@@\int\intkern@\fi                                 %4
 \int}%                                                               %5
\def\multintlimits@{\intop\ifnum\intno@=\z@\intdots@\else\intkern@\fi
 \ifnum\intno@>\tw@\intop\intkern@\fi
 \ifnum\intno@>\thr@@\intop\intkern@\fi\intop}%
\def\intic@{%
    \mathchoice{\hskip.5em}{\hskip.4em}{\hskip.4em}{\hskip.4em}}%
\def\negintic@{\mathchoice
 {\hskip-.5em}{\hskip-.4em}{\hskip-.4em}{\hskip-.4em}}%
\def\ints@@{\iflimtoken@                                              %1
 \def\ints@@@{\iflimits@\negintic@
   \mathop{\intic@\multintlimits@}\limits                             %2
  \else\multint@\nolimits\fi                                          %3
  \eat@}%                                                             %4
 \else                                                                %5
 \def\ints@@@{\iflimits@\negintic@
  \mathop{\intic@\multintlimits@}\limits\else
  \multint@\nolimits\fi}\fi\ints@@@}%
\def\intkern@{\mathchoice{\!\!\!}{\!\!}{\!\!}{\!\!}}%
\def\plaincdots@{\mathinner{\cdotp\cdotp\cdotp}}%
\def\intdots@{\mathchoice{\plaincdots@}%
 {{\cdotp}\mkern1.5mu{\cdotp}\mkern1.5mu{\cdotp}}%
 {{\cdotp}\mkern1mu{\cdotp}\mkern1mu{\cdotp}}%
 {{\cdotp}\mkern1mu{\cdotp}\mkern1mu{\cdotp}}}%
\def\RIfM@{\relax\protect\ifmmode}
\def\text{\RIfM@\expandafter\text@\else\expandafter\mbox\fi}
\let\nfss@text\text
\def\text@#1{\mathchoice
   {\textdef@\displaystyle\f@size{#1}}%
   {\textdef@\textstyle\tf@size{\firstchoice@false #1}}%
   {\textdef@\textstyle\sf@size{\firstchoice@false #1}}%
   {\textdef@\textstyle \ssf@size{\firstchoice@false #1}}%
   \glb@settings}

\def\textdef@#1#2#3{\hbox{{%
                    \everymath{#1}%
                    \let\f@size#2\selectfont
                    #3}}}
\newif\iffirstchoice@
\firstchoice@true
\def\Let@{\relax\iffalse{\fi\let\\=\cr\iffalse}\fi}%
\def\vspace@{\def\vspace##1{\crcr\noalign{\vskip##1\relax}}}%
\def\multilimits@{\bgroup\vspace@\Let@
 \baselineskip\fontdimen10 \scriptfont\tw@
 \advance\baselineskip\fontdimen12 \scriptfont\tw@
 \lineskip\thr@@\fontdimen8 \scriptfont\thr@@
 \lineskiplimit\lineskip
 \vbox\bgroup\ialign\bgroup\hfil$\m@th\scriptstyle{##}$\hfil\crcr}%
\def\Sb{_\multilimits@}%
\def\endSb{\crcr\egroup\egroup\egroup}%
\def\Sp{^\multilimits@}%

\newdimen\ex@
\def\rightarrowfill@#1{$#1\m@th\mathord-\mkern-6mu\cleaders
 \hbox{$#1\mkern-2mu\mathord-\mkern-2mu$}\hfill
 \mkern-6mu\mathord\rightarrow$}%
\def\leftarrowfill@#1{$#1\m@th\mathord\leftarrow\mkern-6mu\cleaders
 \hbox{$#1\mkern-2mu\mathord-\mkern-2mu$}\hfill\mkern-6mu\mathord-$}%
\def\leftrightarrowfill@#1{$#1\m@th\mathord\leftarrow
\mkern-6mu\cleaders
 \hbox{$#1\mkern-2mu\mathord-\mkern-2mu$}\hfill
 \mkern-6mu\mathord\rightarrow$}%
\def\overrightarrow{\mathpalette\overrightarrow@}%
\def\overrightarrow@#1#2{\vbox{\ialign{##\crcr\rightarrowfill@#1\crcr
 \noalign{\kern-\ex@\nointerlineskip}$\m@th\hfil#1#2\hfil$\crcr}}}%

\def\overleftarrow{\mathpalette\overleftarrow@}%
\def\overleftarrow@#1#2{\vbox{\ialign{##\crcr\leftarrowfill@#1\crcr
 \noalign{\kern-\ex@\nointerlineskip}$\m@th\hfil#1#2\hfil$\crcr}}}%
\def\overleftrightarrow{\mathpalette\overleftrightarrow@}%
\def\overleftrightarrow@#1#2{\vbox{\ialign{##\crcr
   \leftrightarrowfill@#1\crcr
 \noalign{\kern-\ex@\nointerlineskip}$\m@th\hfil#1#2\hfil$\crcr}}}%
\def\underrightarrow{\mathpalette\underrightarrow@}%
\def\underrightarrow@#1#2{\vtop{\ialign{##\crcr$\m@th\hfil#1#2\hfil
  $\crcr\noalign{\nointerlineskip}\rightarrowfill@#1\crcr}}}%

\def\underleftarrow{\mathpalette\underleftarrow@}%
\def\underleftarrow@#1#2{\vtop{\ialign{##\crcr$\m@th\hfil#1#2\hfil
  $\crcr\noalign{\nointerlineskip}\leftarrowfill@#1\crcr}}}%
\def\underleftrightarrow{\mathpalette\underleftrightarrow@}%
\def\underleftrightarrow@#1#2{\vtop{\ialign{##\crcr$\m@th
  \hfil#1#2\hfil$\crcr
 \noalign{\nointerlineskip}\leftrightarrowfill@#1\crcr}}}%
\def\qopnamewl@#1{\mathop{\operator@font#1}\nlimits@}
\let\nlimits@\displaylimits
\def\setboxz@h{\setbox\z@\hbox}

\def\varlim@#1#2{\mathop{\vtop{\ialign{##\crcr
 \hfil$#1\m@th\operator@font lim$\hfil\crcr
 \noalign{\nointerlineskip}#2#1\crcr
 \noalign{\nointerlineskip\kern-\ex@}\crcr}}}}

 \def\rightarrowfill@#1{\m@th\setboxz@h{$#1-$}\ht\z@\z@
  $#1\copy\z@\mkern-6mu\cleaders
  \hbox{$#1\mkern-2mu\box\z@\mkern-2mu$}\hfill
  \mkern-6mu\mathord\rightarrow$}
\def\leftarrowfill@#1{\m@th\setboxz@h{$#1-$}\ht\z@\z@
  $#1\mathord\leftarrow\mkern-6mu\cleaders
  \hbox{$#1\mkern-2mu\copy\z@\mkern-2mu$}\hfill
  \mkern-6mu\box\z@$}

\def\projlim{\qopnamewl@{proj\,lim}}
\def\injlim{\qopnamewl@{inj\,lim}}
\def\varinjlim{\mathpalette\varlim@\rightarrowfill@}
\def\varprojlim{\mathpalette\varlim@\leftarrowfill@}
\def\varliminf{\mathpalette\varliminf@{}}
\def\varliminf@#1{\mathop{\underline{\vrule\@depth.2\ex@\@width\z@
   \hbox{$#1\m@th\operator@font lim$}}}}
\def\varlimsup{\mathpalette\varlimsup@{}}
\def\varlimsup@#1{\mathop{\overline
  {\hbox{$#1\m@th\operator@font lim$}}}}

\def\tfrac#1#2{{\textstyle {#1 \over #2}}}%
\def\tint{\mathop{\textstyle \int}}%
\def\tsum{\mathop{\textstyle \sum }}%
\def\tprod{\mathop{\textstyle \prod }}%
\def\stackunder#1#2{\mathrel{\mathop{#2}\limits_{#1}}}%
\begingroup \catcode `|=0 \catcode `[= 1
\catcode`]=2 \catcode `\{=12 \catcode `\}=12
\catcode`\\=12 
|gdef|@alignverbatim#1\end{align}[#1|end[align]]
|gdef|@salignverbatim#1\end{align*}[#1|end[align*]]

|gdef|@alignatverbatim#1\end{alignat}[#1|end[alignat]]
|gdef|@salignatverbatim#1\end{alignat*}[#1|end[alignat*]]

|gdef|@xalignatverbatim#1\end{xalignat}[#1|end[xalignat]]
|gdef|@sxalignatverbatim#1\end{xalignat*}[#1|end[xalignat*]]

|gdef|@gatherverbatim#1\end{gather}[#1|end[gather]]
|gdef|@sgatherverbatim#1\end{gather*}[#1|end[gather*]]

|gdef|@gatherverbatim#1\end{gather}[#1|end[gather]]
|gdef|@sgatherverbatim#1\end{gather*}[#1|end[gather*]]

|gdef|@multilineverbatim#1\end{multiline}[#1|end[multiline]]
|gdef|@smultilineverbatim#1\end{multiline*}[#1|end[multiline*]]

|gdef|@arraxverbatim#1\end{arrax}[#1|end[arrax]]
|gdef|@sarraxverbatim#1\end{arrax*}[#1|end[arrax*]]

|gdef|@tabulaxverbatim#1\end{tabulax}[#1|end[tabulax]]
|gdef|@stabulaxverbatim#1\end{tabulax*}[#1|end[tabulax*]]

|endgroup

\def\align{\@verbatim \frenchspacing\@vobeyspaces \@alignverbatim
You are using the "align" environment in a style in which it is not defined.}

\@namedef{align*}{\@verbatim\@salignverbatim
You are using the "align*" environment in a style in which it is not defined.}
\expandafter\let\csname endalign*\endcsname =\endtrivlist

\def\alignat{\@verbatim \frenchspacing\@vobeyspaces \@alignatverbatim
You are using the "alignat" environment in a style in which it is not defined.}

\@namedef{alignat*}{\@verbatim\@salignatverbatim
You are using the "alignat*" environment in a style in which it is not defined.}
\expandafter\let\csname endalignat*\endcsname =\endtrivlist

\def\xalignat{\@verbatim \frenchspacing\@vobeyspaces \@xalignatverbatim
You are using the "xalignat" environment in a style in which it is not defined.}

\@namedef{xalignat*}{\@verbatim\@sxalignatverbatim
You are using the "xalignat*" environment in a style in which it is not defined.}
\expandafter\let\csname endxalignat*\endcsname =\endtrivlist

\def\gather{\@verbatim \frenchspacing\@vobeyspaces \@gatherverbatim
You are using the "gather" environment in a style in which it is not defined.}

\@namedef{gather*}{\@verbatim\@sgatherverbatim
You are using the "gather*" environment in a style in which it is not defined.}
\expandafter\let\csname endgather*\endcsname =\endtrivlist

\def\multiline{\@verbatim \frenchspacing\@vobeyspaces \@multilineverbatim
You are using the "multiline" environment in a style in which it is not defined.}

\@namedef{multiline*}{\@verbatim\@smultilineverbatim
You are using the "multiline*" environment in a style in which it is not defined.}
\expandafter\let\csname endmultiline*\endcsname =\endtrivlist

\def\arrax{\@verbatim \frenchspacing\@vobeyspaces \@arraxverbatim
You are using a type of "array" construct that is only allowed in AmS-LaTeX.}

\def\tabulax{\@verbatim \frenchspacing\@vobeyspaces \@tabulaxverbatim
You are using a type of "tabular" construct that is only allowed in AmS-LaTeX.}

\@namedef{arrax*}{\@verbatim\@sarraxverbatim
You are using a type of "array*" construct that is only allowed in AmS-LaTeX.}
\expandafter\let\csname endarrax*\endcsname =\endtrivlist

\@namedef{tabulax*}{\@verbatim\@stabulaxverbatim
You are using a type of "tabular*" construct that is only allowed in AmS-LaTeX.}
\expandafter\let\csname endtabulax*\endcsname =\endtrivlist

\def\@@eqncr{\let\@tempa\relax
    \ifcase\@eqcnt \def\@tempa{& & &}\or \def\@tempa{& &}%
      \else \def\@tempa{&}\fi
     \@tempa
     \if@eqnsw
        \iftag@
           \@taggnum
        \else
           \@eqnnum\stepcounter{equation}%
        \fi
     \fi
     \global\tag@false
     \global\@eqnswtrue
     \global\@eqcnt\z@\cr}

 \def\endequation{%
     \ifmmode\ifinner % FLEQN hack
      \iftag@
        \addtocounter{equation}{-1} % undo the increment made in the begin part
        $\hfil
           \displaywidth\linewidth\@taggnum\egroup \endtrivlist
        \global\tag@false
        \global\@ignoretrue   
      \else
        $\hfil
           \displaywidth\linewidth\@eqnnum\egroup \endtrivlist
        \global\tag@false
        \global\@ignoretrue 
      \fi
     \else   
      \iftag@
        \addtocounter{equation}{-1} % undo the increment made in the begin part
        \eqno \hbox{\@taggnum}
        \global\tag@false%
        $$\global\@ignoretrue
      \else
        \eqno \hbox{\@eqnnum}% $$ BRACE MATCHING HACK
        $$\global\@ignoretrue
      \fi
     \fi\fi
 } 

 \newif\iftag@ \tag@false
 
 \def\tag{\@ifnextchar*{\@tagstar}{\@tag}}
 \def\@tag#1{%
     \global\tag@true
     \global\def\@taggnum{(#1)}}
 \def\@tagstar*#1{%
     \global\tag@true
     \global\def\@taggnum{#1}%  
}

\makeatother
\begin{document}

\TeXButton{flushright}
{\begin{flushright}
\setlength{\baselineskip}{3ex}
\#HUTP-98/A023\\ 5/98\end{flushright}
}

\TeXButton{TeX field}
{
\begin{center}
{\Large \textbf{{Introduction to spherical field theory} }{\normalsize \textrm{\\ \vspace{12pt} 
Dean Lee\footnote{Supported by
the National Science Foundation under Grant \#PHY-9218167 and the Fannie and
John Hertz Foundation.}
\\ Harvard University\\ Cambridge, MA 02138\\ \vspace{24pt} 
\small
\parbox{360pt}{Spherical field theory is a new non-perturbative method for studying quantum field 
theories. It uses the spherical partial wave expansion to reduce a 
general $d$-dimensional Euclidean field theory into a set of coupled 
one-dimensional systems. The coupled one-dimensional systems are then converted to partial 
differential equations and solved numerically. We demonstrate the methods of spherical 
field theory by analyzing Euclidean $\phi ^4$ theory in two dimensions.
[PACS numbers: 11.10.Kk, 11.15Tk, 11.30Qc]}
\normalsize
 \vspace{10pt} }}}
\end{center}
}\TeXButton{setcounter}{\setcounter{footnote}{0}}

\section{Overview}

We present a new non-perturbative method for studying quantum field
theories. We describe a general procedure for converting problems in quantum
field theory into initial-value problems described by partial differential
equations. Our derivation of the method starts with the functional integral
for a $d$-dimensional Euclidean field theory. Making use of spherical $O(d)$%
-symmetry, we decompose the fields as a linear combination of spherical
partial waves. Re-interpreting the radial part of each partial wave as a
distinct field in a new one-dimensional field theory, we rewrite the
functional integral as a time-dependent Schr\"{o}dinger equation, where the
radial distance in the original theory serves as the time parameter.
Although the resulting Schr\"{o}dinger equation contains an infinite set of
interacting partial waves, we find that high spin partial waves decouple if
the original field theory is renormalized to remove ultraviolet divergences.
Neglecting higher spin partial waves and estimating the corresponding error,
we solve the remaining partial differential equation using standard methods.

The organization of this paper is as follows. We begin with a short
discussion of one-dimensional systems and the connection between spherical
field theory and coupled one-dimensional systems. Next we illustrate the
method by analyzing normal-ordered Euclidean $\phi ^4$ theory in two
dimensions. We derive the spherical Feynman rules for this theory, compute
the two-loop self-energy, and compare the result with standard perturbation
theory. We then consider non-perturbative values of the coupling constant
and evaluate the self-energy numerically using finite-difference methods and
diffusion Monte Carlo. Since our paper is intended as an introduction to the
methods of spherical field theory rather than a detailed analysis of
Euclidean $\phi ^4$ theory, we have not engaged in extensive numerical
computations for our analysis here. Nevertheless, our rather modest
calculations show clear evidence of spontaneously broken $\phi \rightarrow
-\phi $ reflection symmetry at the critical coupling value 
\begin{equation}
\frac \lambda {4!(2\pi )\mu ^2}\simeq 0.4,  \label{bo}
\end{equation}
where $\mu $ is the bare mass and $\lambda $ is defined by the quartic
coupling $\frac \lambda {4!}\phi ^4$. Our value of the critical coupling is
in close agreement with the results of a recent lattice study \cite{wi}. The
methods we present in this paper are easily generalized to any $d$%
-dimensional Euclidean field theory involving any number of scalar fields.
The extension to fermionic and gauge systems is the subject of current work.

\section{Spherical fields}

We start with the Feynman-Kac formula, which connects the Schr\"{o}dinger
and path integral representations of (Euclideanized) quantum mechanics, 
\begin{eqnarray}
&&\left\langle x_F\right| T\exp \left[ -\tint\nolimits_{t_I}^{t_F}dt\,H_{%
\mathcal{J}}\right] \left| x_I\right\rangle   \label{rm} \\
&\propto &\int\limits_{\Sb \phi (t_I)=x_I \\ \phi (t_F)=x_F \endSb }\mathcal{%
D}\phi \exp \left\{ -\tint\nolimits_{t_I}^{t_F}\,dt\left[ \tfrac 12\left( 
\tfrac{\partial \phi }{\partial t}\right) ^2+V(\phi )-\mathcal{J}\phi
\right] \right\} ,  \nonumber
\end{eqnarray}
where 
\begin{eqnarray}
H_{\mathcal{J}}(t) &=&-\tfrac 12\tfrac{d^2}{dq^2}+V(q)-\mathcal{J}q
\label{e} \\
&=&H_0-\mathcal{J}q  \nonumber
\end{eqnarray}
and $\left| x_I\right\rangle ,$ $\left| x_F\right\rangle $ are position
eigenstates centered at $x_I,$ $x_F.$ The generating functional $Z[\mathcal{J%
}]$ is proportional to the ground state persistence amplitude, 
\begin{equation}
Z[\mathcal{J}]\propto \left\langle 0\right| T\exp \left[
-\tint\nolimits_{-\infty }^\infty dt\,H_{\mathcal{J}}\right] \left|
0\right\rangle ,  \label{zx}
\end{equation}
where $\left| 0\right\rangle $ is the ground state of $H_0.$ For $t_1>t_2,$%
\begin{equation}
\left. \tfrac 1{Z[0]}\tfrac{\delta ^2Z[\mathcal{J}]}{\delta \mathcal{J}%
(t_1)\delta \mathcal{J}(t_2)}\right| _{\mathcal{J}=0}=\tfrac{\left\langle
0\right| T\exp \left[ -\int_{t_1}^\infty dt\,H_0\right] \,q\,T\exp \left[
-\int_{t_2}^{t_1}dt\,H_0\right] \,q\,T\exp \left[ -\int_{-\infty
}^{t_2}dt\,H_0\right] \left| 0\right\rangle }{\left\langle 0\right| T\exp
\left[ -\int_{-\infty }^\infty dt\,H_0\right] \left| 0\right\rangle }.
\label{en1}
\end{equation}
The Feynman-Kac formula can be generalized to include more general
time-dependent interactions, 
\begin{eqnarray}
&&\left\langle x_F\right| T\exp \left[
-\tint\nolimits_{t_I}^{t_F}dt\,H(t)\right] \left| x_I\right\rangle 
\label{bz} \\
&\propto &\int\limits_{\Sb \phi (t_I)=x_I \\ \phi (t_F)=x_F \endSb }\mathcal{%
D}\phi \exp \left\{ -\tint\nolimits_{t_I}^{t_F}\,dt\left[ \tfrac
1{2a(t)}\left( \tfrac{\partial \phi }{\partial t}\right) ^2+V(\phi
,t)\right] \right\} ,  \nonumber
\end{eqnarray}
where 
\begin{equation}
H(t)=-\tfrac{a(t)}2\tfrac{d^2}{dq^2}+V(q,t).  \label{vi}
\end{equation}

The central idea of spherical field theory is to treat a general $d$%
-dimensional field theory as a set of coupled one-dimensional systems. To
demonstrate, we consider Euclidean $\phi ^4$ theory in two dimensions, 
\begin{equation}
Z[\mathcal{J}]=\int \mathcal{D}\phi \exp \left\{ -\tint d^2x\left[ \tfrac
12\sum_i(\partial _i\phi )^2+\tfrac{\mu ^2}2\phi ^2+\tfrac \lambda {4!}\phi
^4-\mathcal{J}\phi \right] \right\} .  \label{bw}
\end{equation}
Decomposing $\phi $ and $\mathcal{J}$ into partial waves, we have

\begin{equation}
\phi (t\cos \theta ,t\sin \theta )=\sqrt{\tfrac 1{2\pi }}\tsum_{n=0,\pm
1,...}\phi _n(t)\,e^{in\theta },  \label{nx}
\end{equation}
\begin{equation}
\mathcal{J}(t\cos \theta ,t\sin \theta )=\sqrt{\tfrac 1{2\pi }}%
\tsum_{n=0,\pm 1,...}\mathcal{J}_n(t)\,e^{in\theta }.  \label{mr}
\end{equation}
The decomposition of $\phi $ into partial waves is a linear invertible
transformation. The Jacobian of the transformation (tempered with suitable
ultraviolet and infrared regulation) is, 
\begin{equation}
\text{det}\left[ \tfrac{\delta \phi (t^{\prime }\cos \theta ,t^{\prime }\sin
\theta )}{\delta \phi _n(t)}\right] =\text{det}\left[ \sqrt{\tfrac 1{2\pi }}%
\delta (t-t^{\prime })\,e^{in\theta }\right] .  \label{me}
\end{equation}
This Jacobian is independent of $\phi _n(t)$, and we change integration
variables to write $Z[\mathcal{J}]$ as 
\begin{equation}
b\int \tprod_{n=0,\pm 1,...}\mathcal{D}\phi _n\exp \left\{
-\tint\nolimits_0^\infty dt\left[ 
\begin{array}{c}
\sum\limits_{n=0,\pm 1,...}\frac t2\frac{\partial \phi _{-n}}{\partial t}%
\frac{\partial \phi _n}{\partial t}+\frac{\mu ^2t^2+n_{}^2}{2t}\phi
_{-n}\phi _n-t\mathcal{J}_{-n}\phi _n \\ 
+\frac{\lambda t}{4!2\pi }\sum\limits\Sb n_1,n_2,n_3  \\ =0,\pm 1,... 
\endSb \phi _{n_1}\phi _{n_2}\phi _{n_3}\phi _{-n_1-n_2-n_3}
\end{array}
\right] \right\} ,  \label{ty}
\end{equation}
where $b$ is an overall constant whose value is irrelevant for the purposes
of our discussion. From (\ref{bz}) and (\ref{vi}), this functional integral
corresponds with the Schr\"{o}dinger time evolution generator 
\begin{eqnarray}
H_{\mathcal{J}} &=&\sum\limits_{n=0,\pm 1,...}-\tfrac 1{2t}\tfrac \partial
{\partial q_{-n}}\tfrac \partial {\partial q_n}+\tfrac{\mu ^2t^2+n_{}^2}{2t}%
q_{-n}q_n-t\mathcal{J}_{-n}q_n  \label{bn} \\
&&+\tfrac{\lambda t}{4!2\pi }\sum\limits\Sb n_1,n_2,n_3  \\ =0,\pm 1,... 
\endSb q_{n_1}q_{n_2}q_{n_3}q_{-n_1-n_2-n_3}.  \nonumber
\end{eqnarray}
In order to define $Z[\mathcal{J}]$ in the Schr\"{o}dinger language, we
first restrict the $t$-integral in (\ref{ty}) to a finite interval, $t_I\leq
t\leq t_F.$ We then have 
\begin{equation}
Z_{[t_I,t_F]}[\mathcal{J}]\propto \int \tprod_{n=0,\pm
1,...}dx_ndx_n^{\prime }\,\left\langle \left\{ x_n^{\prime }\right\}
_{n=0,\pm 1,\cdots }\right| T\exp \left[ -\tint\nolimits_{t_I}^{t_F}dt\,H_{%
\mathcal{J}}(t)\right] \left| \left\{ x_n\right\} _{n=0,\pm 1,\cdots
}\right\rangle ,  \label{iv}
\end{equation}
where 
\begin{equation}
q_i\left| \left\{ x_n\right\} _{n=0,\pm 1,\cdots }\right\rangle =x_i\left|
\left\{ x_n\right\} _{n=0,\pm 1,\cdots }\right\rangle .  \label{eb}
\end{equation}
We recover $Z[\mathcal{J}]$ by taking the limits $t_I\rightarrow 0$, $%
t_F\rightarrow \infty $. The details of the quantum state at the boundary $%
t_F$ is rather unimportant since only the ground state projection survives
as $t_F\rightarrow \infty $, similar to what happens in (\ref{zx}). In
contrast, the $q_0$ dependence of the quantum state at $t_I=0$ is
significant.\footnote{%
For $n\neq 0$ the $q_n$ dependence is again unimportant and depends only on
the ground state projection.} The proper initial state is 
\begin{equation}
\left| a\right\rangle \equiv \int \tprod_{n=0,\pm 1,...}dx_n\left| \left\{
x_n\right\} _{n=0,\pm 1,\cdots }\right\rangle ,  \label{vb}
\end{equation}
a state whose wavefunction is constant in $q$-space.

\section{Correlators}

The generating functional, written as a power series, is

\begin{equation}
\tfrac{Z[\mathcal{J}]}{Z[0]}=1+\tfrac 12\tint \tfrac{d^2\vec{p}d^2\vec{x}d^2%
\vec{y}}{(2\pi )^2}e^{-i\vec{p}\cdot \vec{x}}e^{i\vec{p}\cdot \vec{y}}%
\mathcal{J}(\vec{x})\mathcal{J}(\vec{y})f(\vec{p}^2)+\cdots ,  \label{mv}
\end{equation}
where $f(\vec{k}^2)$ is the $\phi $ propagator, 
\begin{equation}
f(\vec{k}^2)=\tint d^2\vec{x}\,e^{i\vec{k}\cdot \vec{x}}\left\langle
0\right| \phi (\vec{x})\phi (0)\left| 0\right\rangle .  \label{dr}
\end{equation}
As a result of angular momentum conservation, the term quadratic in $%
\mathcal{J}$ can be decomposed as a sum over terms proportional to $\mathcal{%
J}_{-n}\mathcal{J}_n,$%
\begin{equation}
\tfrac 12\sum\limits_{n=0,\pm 1,...}\tint dkdt_xdt_y\,t_xt_yk\mathcal{J}%
_{-n}(t_x)\mathcal{J}_n(t_y)J_{\left| n\right| }(kt_x)J_{\left| n\right|
}(kt_y)\,f(k^2),  \label{ke}
\end{equation}
where $J_i$ is the $i^{\text{th}}$ order Bessel function of the first kind.
The two-point correlator for the $n$-th partial wave is 
\begin{equation}
\left\langle 0\right| \phi _{-n}(t_1)\phi _n(t_2)\left| 0\right\rangle
=\left. \tfrac 1{Z[0]}\tfrac{\delta ^2Z[\mathcal{J}]}{t_1\delta \mathcal{J}%
_n(t_1)t_2\delta \mathcal{J}_{-n}(t_2)}\right| _{\mathcal{J}=0}=\tint
dk\,k\,J_{\left| n\right| }(kt_1)J_{\left| n\right| }(kt_2)f(k^2).
\label{cu}
\end{equation}
We will use this result in the next section when discussing perturbative
spherical field theory. For the special case $n=0$ and $t_2=0$ we have 
\begin{eqnarray}
\left\langle 0\right| \phi _0(t_1)\phi _0(0)\left| 0\right\rangle &=&\tint
dk\,k\,J_0(kt_1)f(k^2)=\tint \tfrac{d^2\vec{p}}{2\pi }e^{-i\vec{p}\cdot \vec{%
x}}f(\vec{p}^2)  \label{b} \\
&=&2\pi \,\left\langle 0\right| \phi (\vec{r})\phi (0)\left| 0\right\rangle ,
\nonumber
\end{eqnarray}
when $\left| \vec{r}\right| =t_1.$ The two-point $\phi $ correlator is
proportional to the two-point $\phi _0$ correlator with one $\phi _0$
centered at the origin. We can compute the two-point $\phi _0$ correlator
using the Schr\"{o}dinger formalism, 
\begin{equation}
\left\langle 0\right| \phi _0(t_1)\phi _0(0)\left| 0\right\rangle =\tfrac{%
\left\langle a\right| T\exp \left[ -\int_{t_1}^\infty dt\,H_0\right]
\,q_0\,T\exp \left[ -\int_0^{t_1}dt\,H_0\right] \,q_0\,\left| a\right\rangle 
}{\left\langle a\right| T\exp \left[ -\int_0^\infty dt\,H_0\right] \left|
a\right\rangle }.  \label{mc}
\end{equation}
More general correlation functions can be evaluated by partial wave
expansion. For example the four-point $\phi $ correlator is given by 
\begin{equation}
\left\langle 0\right| \phi (\vec{r}_1)\phi (\vec{r}_2)\phi (\vec{r}_3)\phi
(0)\left| 0\right\rangle =\sum_{\Sb n_1,n_2,n_3  \\ =0,\pm 1,...  \endSb }%
\tfrac{e^{i(n_1\theta _1+n_2\theta _2+n_3\theta _3)}}{(2\pi )^2}\left\langle
0\right| \phi _{n_1}(t_1)\phi _{n_2}(t_2)\phi _{n_3}(t_3)\phi _0(0)\left|
0\right\rangle  \label{cx}
\end{equation}
where 
\begin{equation}
\vec{r}_i=(t_i\cos \theta _i,t_i\sin \theta _i),\qquad i=1,2,3.  \label{hs}
\end{equation}
In the Schr\"{o}dinger formalism, we can rewrite 
\begin{equation}
\left\langle 0\right| \phi _{n_1}(t_1)\phi _{n_2}(t_2)\phi _{n_3}(t_3)\phi
_0(0)\left| 0\right\rangle  \label{hx}
\end{equation}
as (for $t_1\geq t_2\geq t_3)$, 
\begin{equation}
\tfrac{\left\langle a\right| T\exp \left[ -\int_{t_1}^\infty dt\,H_0\right]
q_{n_1}\,T\exp \left[ -\int_{t_2}^{t_1}dt\,H_0\right] q_{n_2}T\exp \left[
-\int_{t_3}^{t_2}dt\,H_0\right] q_{n_3}\,T\exp \left[
-\int_0^{t_3}dt\,H_0\right] \,q_0\,\left| a\right\rangle }{\left\langle
a\right| T\exp \left[ -\int_0^\infty dt\,H_0\right] \left| a\right\rangle }.
\label{re}
\end{equation}

\section{Perturbative spherical field theory}

For free field theory we have $f(\vec{k}^2)=\tfrac 1{\vec{k}^2+\mu ^2}.$
From (\ref{cu}), 
\begin{equation}
\left\langle 0\right| \phi _{-n}(t_1)\phi _n(t_2)\left| 0\right\rangle
=\theta (t_1-t_2)K_{\left| n\right| }(\mu t_1)I_{\left| n\right| }(\mu
t_2)+\theta (t_2-t_1)K_{\left| n\right| }(\mu t_2)I_{\left| n\right| }(\mu
t_1),  \label{ne}
\end{equation}
where $I_i$ and $K_i$ are the $i^{\text{th}}$ order modified Bessel
functions of the first and second kind respectively. In particular, we find 
\begin{equation}
\left\langle 0\right| \phi _0(t)\phi _0(0)\left| 0\right\rangle =K_0(\mu t).
\label{bvr}
\end{equation}

We can use (\ref{ne}) to generate the spherical version of Feynman diagrams.
As an example, the two-loop process shown in Figure 1 gives us 
\begin{equation}
\int dt_1dt_2\tfrac{\lambda ^2t_1t_2}{(2\pi )^2}\left[ 
\begin{array}{c}
\theta (t-t_1)K_0(\mu t)I_0(\mu t_1) \\ 
+\theta (t_1-t)K_0(\mu t_1)I_0(\mu t)
\end{array}
\right] \left[ 
\begin{array}{c}
\theta (t_1-t_2)K_{1,2,3}(\mu t_1)I_{1,2,3}(\mu t_2) \\ 
+\theta (t_2-t_1)K_{1,2,3}(\mu t_2)I_{1,2,3}(\mu t_1)
\end{array}
\right] K_0(\mu t_2),  \label{ye}
\end{equation}
where 
\begin{eqnarray}
I_{1,2,3}(x) &=&I_1(x)I_2(x)I_3(x),  \label{nn} \\
K_{1,2,3}(x) &=&K_1(x)K_2(x)K_3(x).  \nonumber
\end{eqnarray}
This diagram contributes to the two-point $\phi _0$ correlator at order $%
\lambda ^2$. Let $\Delta _{2,J_{\max }}(t)$ be the sum of all such $%
O(\lambda ^2)$ diagrams for the $\phi _0$ correlator, with the restriction
that each internal line correspond with some $\phi _n$ such that $\left|
n\right| \leq J_{\max }$. In view of relation (\ref{b}) we expect that as $%
J_{\max }\rightarrow \infty $ we should recover the regular two-loop
perturbative result, 
\begin{equation}
\lim_{J_{\max }\rightarrow \infty }\tfrac 1{2\pi }\int dt\,d\theta
\,te^{itk\cos \theta }\Delta _{2,J_{\max }}(t)\stackunder{}{=}\tfrac{\lambda
^2}{96\pi ^2(k^2+\mu ^2)^2}\int_0^1\int_0^{1-\alpha _1}\tfrac{d\alpha
_2d\alpha _1\delta (\alpha _1+\alpha _2+\alpha _3-1)}{\alpha _1\alpha
_2\alpha _3k^2+(\alpha _1\alpha _2+\alpha _2\alpha _3+\alpha _3\alpha _1)\mu
^2}.  \label{bb}
\end{equation}
In Figure 2 we have plotted 
\begin{equation}
s_{J_{\max }}(k^2)\equiv \tfrac{48\pi (k^2+\mu ^2)^2}{\lambda ^2}\int
dt\,d\theta \,te^{itk\cos \theta }\Delta _{2,J_{\max }}(t),\qquad J_{\max
}=0,1,2,3,  \label{hh}
\end{equation}
and 
\begin{equation}
s_\infty (k^2)\equiv \int_0^1\int_0^{1-\alpha _1}\tfrac{d\alpha _2d\alpha _1%
}{\alpha _1\alpha _2\alpha _3k^2+(\alpha _1\alpha _2+\alpha _2\alpha
_3+\alpha _3\alpha _1)\mu ^2},  \label{cd}
\end{equation}
in units where $\mu =1.$ It appears that $s_{J_{\max }}(k^2)$ does in fact
converge towards $s_\infty (k^2).$ We discuss this and related issues in the
next section.

\section{High spin decoupling}

Partial waves with high spin correspond with high tangential momentum modes.
We can see this explicitly in the coefficient of $\phi _{-n}\phi _n$ in (\ref
{ty}), 
\begin{equation}
\tfrac{\mu ^2t^2+n^2}{2t},  \label{nb}
\end{equation}
where the $n^2$ dependence is generated by the centrifugal part of the
kinetic term 
\begin{equation}
\tfrac 12\sum_i(\partial _i\phi )^2.  \label{bs}
\end{equation}
Therefore if our field theory is renormalized to remove ultraviolet
divergences, the contribution of high spin partial waves will decouple in
the same manner as other high momentum effects such as dependence on an
explicit cutoff mass scale. The error in neglected partial waves with spin
larger than some value $J_{\max }$ is roughly equivalent to the error in
using a cutoff mass $\Lambda _{\text{eff}}$ where 
\begin{equation}
\Lambda _{\text{eff}}^2\sim \tfrac{J_{\max }^2}{t^2},  \label{xc}
\end{equation}
and $t$ is the characteristic radius for the process being studied.

In two-dimensional $\phi ^4$ theory, renormalization can be done simply by
normal ordering the interaction. In the context of spherical field, normal
ordering corresponds with explicitly subtracting the contribution of
diagrams of the type shown in Figure 3. The amplitude for this diagram is
proportional to 
\begin{equation}
\lambda tK_{\left| n\right| }(\mu t)I_{\left| n\right| }(\mu t).  \label{cj2}
\end{equation}
In the Schr\"{o}dinger language this corresponds with a counterterm
proportional to 
\begin{equation}
\lambda tK_{\left| n\right| }(\mu t)I_{\left| n\right| }(\mu t)q_{-i}q_i.
\label{dh}
\end{equation}

Let us now return to the two-loop perturbative calculation. In presence of
an explicit cutoff mass $\Lambda _{\text{eff}}$, the $\Lambda _{\text{eff}}$%
-regulated calculation for the two-loop result $s_\infty (k^2)$ includes a
correction of size $O(\Lambda _{\text{eff}}^{-2})$. From relation (\ref{xc}%
), we expect 
\begin{equation}
s_\infty (k^2)-s_{J_{\max }}(k^2)\sim O(J_{\max }^{-2}).  \label{if}
\end{equation}

\section{Non-perturbative self-energy}

Let us define $c=\tfrac \lambda {4!(2\pi )}$ and $\Delta (t)$ to be the
difference between the two-point $\phi _0$ correlator for general $c$ and $%
c=0,$ 
\begin{equation}
\Delta (t)=\left\langle 0\right| \phi _0(t)\phi _0(0)\left| 0\right\rangle
_c-\left\langle 0\right| \phi _0(t)\phi _0(0)\left| 0\right\rangle _{c=0}.
\label{bi}
\end{equation}
We can compute $\Delta (t)$ using (\ref{bn}) and (\ref{mc}) along with
normal-ordering counterterms described in the previous section. For $c=0.20$
we have computed $\Delta (t)$ for $J_{\max }=0$ and $J_{\max }=1.\ $For $%
J_{\max }=0$ the Schr\"{o}dinger time evolution generator is 
\begin{equation}
-\tfrac 1{2t}\tfrac{\partial ^2}{\partial q_0^2}+\tfrac{\mu ^2t}%
2q_0^2+ctq_0^4-6ctK_0(\mu t)I_0(\mu t)q_0^2.  \label{be}
\end{equation}
For $J_{\max }=1$ the time evolution generator is 
\begin{eqnarray}
&&-\tfrac 1{2t}\left[ \tfrac{\partial ^2}{\partial q_0^2}+\tfrac{\partial ^2%
}{\partial q_{s1}^2}+\tfrac{\partial ^2}{\partial q_{a1}^2}\right] +\tfrac{%
\mu ^2t}2q_0^2+\tfrac{\mu ^2t^2+1}{2t}(q_{s1}^2+q_{a1}^2)  \label{en2} \\
&&+ct\left[ q_0^4+\tfrac
32(q_{s1}^2+q_{a1}^2)^2+6q_0^2(q_{s1}^2+q_{a1}^2)\right]  \nonumber \\
&&-6ct\left[ K_0(\mu t)I_0(\mu t)+2K_1(\mu t)I_1(\mu t)\right] \left[
q_0^2+q_{s1}^2+q_{a1}^2\right] ,  \nonumber
\end{eqnarray}
where 
\begin{equation}
q_{s1}=\tfrac{q_1+q_{-1}}{\sqrt{2}},\;q_{a1}=\tfrac{q_1-q_{-1}}{i\sqrt{2}}.
\label{zt1}
\end{equation}
For each value of $J_{\max }$ we did the computation two ways, once using
the method of finite differences and once using diffusion Monte Carlo. The
results are shown in Figure 4, where we have used the abbreviations fd for
finite difference method and mc for Monte Carlo. This plot should be
regarded as a measure of the accuracy of the Monte Carlo algorithm. The
error of the finite difference algorithm is likely negligible as we were
able to reproduce the two-point $\phi _0$ correlator for free field theory
to an accuracy of 0.1\%.

Let $\Pi (k^2)$ denote the $\phi $ self-energy, defined such that the full $%
\phi $ propagator is

\begin{equation}
f(k^2)=\tfrac 1{k^2+\mu ^2+\Pi (k^2)}.  \label{vy}
\end{equation}
We have calculated $-\Pi (k^2)$ for $c=0.05,0.10,\cdots ,0.40$ and $J_{\max
}=2$. For $J_{\max }=2$ the time evolution generator is 
\begin{eqnarray}
&&-\tfrac 1{2t}\left[ \tfrac{\partial ^2}{\partial q_0^2}+\tfrac{\partial ^2%
}{\partial q_{s1}^2}+\tfrac{\partial ^2}{\partial q_{a1}^2}+\tfrac{\partial
^2}{\partial q_{s2}^2}+\tfrac{\partial ^2}{\partial q_{a2}^2}\right]
\label{zs} \\
&&+\tfrac{\mu ^2t}2q_0^2+\tfrac{\mu ^2t^2+1}{2t}(q_{s1}^2+q_{a1}^2)+\tfrac{%
\mu ^2t^2+4}{2t}(q_{s2}^2+q_{a2}^2)  \nonumber \\
&&+ct\left[ 
\begin{array}{c}
q_0^4+\tfrac 32(q_{s1}^2+q_{a1}^2)^2+\tfrac
32(q_{s2}^2+q_{a2}^2)^2+6q_0^2(q_{s1}^2+q_{a1}^2+q_{s2}^2+q_{a2}^2) \\ 
+6\sqrt{2}%
q_0(q_{s1}^2q_{s2}-q_{a1}^2q_{s2}+2q_{a1}q_{s1}q_{a2})+6(q_{s1}^2+q_{a1}^2)(q_{s2}^2+q_{a2}^2)
\end{array}
\right]  \nonumber \\
&&-6ct\left[ K_0(\mu t)I_0(\mu t)+2K_1(\mu t)I_1(\mu t)+2K_2(\mu t)I_2(\mu
t)\right] \cdot \left[ 
\begin{array}{c}
q_0^2+q_{s1}^2+q_{a1}^2 \\ 
+q_{s2}^2+q_{a2}^2
\end{array}
\right]  \nonumber
\end{eqnarray}
where 
\begin{equation}
q_{s2}=\tfrac{q_2+q_{-2}}{\sqrt{2}},\;q_{a2}=\tfrac{q_2-q_{-2}}{i\sqrt{2}}.
\label{zt2}
\end{equation}
Our results were calculated using diffusion Monte Carlo and are shown in
Figure 5. The physical squared-mass $\mu _{\text{phys}}^2$ can be found by
extrapolating $-\Pi (k^2)$ to its intersection point with the line $k^2+1$
(in units where the bare mass $\mu =1),$ 
\begin{equation}
-\Pi (-\mu _{\text{phys}}^2)=-\mu _{\text{phys}}^2+1.  \label{zt3}
\end{equation}
One fact that is useful in performing this extrapolation is 
\begin{equation}
\left. -\tfrac{d\Pi }{dk^2}\right| _{k^2=-\mu _{\text{phys}}^2}\leq 0,
\label{cj1}
\end{equation}
since the residue of the pole at $-\mu _{\text{phys}}^2$ must be less than
or equal to $1$.\footnote{%
This can be proved by spectral decomposition of the two-point $\phi $
correlator.} As Figure 5 shows (and also visible in the plots of $s_{J_{\max
}}(k^2)$ in Figure 2), the slope of $-\Pi (k^2)$ appears to have the wrong
sign at small values of $k^2$. The reason for this is that high spin partial
waves cannot be neglected for $t\sim O(J_{\max }^{})$, or equivalently $%
k^2\sim O(J_{\max }^{-2})$. It is therefore necessary to extrapolate $-\Pi
(k^2)$ from values of $k^2$ where this breakdown does not occur.

Extrapolating $-\Pi (k^2)$ through the small $k^2$ region is somewhat
difficult for $J_{\max }=2$, and we could achieve greater accuracy by
choosing a higher value for $J_{\max }$. But, as mentioned before, our main
goal here is to introduce to the methods of spherical field theory. A more
thorough and accurate calculation of this and other phenomena will be done
in the future. For now the extrapolation method we choose is simply to
extend the maximum value of $-\Pi (k^2)$ to all $k^2$ to the left of the
maximum, as shown by the horizontal dashed lines in Figure 5. With this
simple procedure we find the following values of $\mu _{\text{phys}}^2$.

\begin{equation}
\begin{array}{ccc}
c &  & \mu _{\text{phys}}^2 \\ 
0 &  & 1 \\ 
0.05 &  & 0.95 \\ 
0.10 &  & 0.8 \\ 
0.15 &  & 0.65 \\ 
0.20 &  & 0.5 \\ 
0.25 &  & 0.35 \\ 
0.30 &  & 0.2 \\ 
0.35 &  & 0.1 \\ 
0.40 &  & 0.0
\end{array}
\label{xz}
\end{equation}
For $c$ near $0.4$ the physical mass vanishes. As discussed in the
literature, \cite{gj} and \cite{gjs}$,$ this is associated with spontaneous
breaking of $\phi \rightarrow -\phi $ reflection symmetry. Despite using a
small value for $J_{\max }$ and a minimal amount of computer labor, we find
that our value for the critical coupling is in agreement with a recent
lattice calculation, \cite{wi}, which found the critical coupling at $%
c=0.407 $.

\section{Summary}

Spherical field theory is a new approach to studying quantum field theory.
It treats a general $d$-dimensional field theory as a system of coupled
one-dimensional theories, which are then solved using partial differential
equations. In this paper we introduced the methods of spherical field theory
using the example of Euclidean $\phi ^4$ theory in two dimensions. We first
derived the spherical Feynman rules, showed that high spin partial waves
decouple, and then, using a spin cutoff $J_{\max }=2,$ computed the $\phi $
self-energy for non-perturbative values of the coupling. Our results show
clear evidence of a phase transition near $c=0.4$. The methods used here are
easily generalized to any $d$-dimensional Euclidean field theory involving
scalar fields. The extension to fermionic and gauge systems is currently
being studied.

We expect spherical field theory to have many applications in the areas of
particle theory and statistical physics. It offers an efficient and flexible
way to calculate non-perturbative quantities in field theory. For simple
systems (small $J_{\max }$) either the method of finite differences or Monte
Carlo simulation can be used. For larger systems only Monte Carlo methods
are feasible, but here again we have several qualitatively different
options. In our computation of the $\phi $ self-energy we used diffusion
Monte Carlo to model the time evolution equation. Another option, though, is
to compute the one-dimensional path integral shown in (\ref{ty}) using
standard lattice field methods. The ability to calculate quantities in
several different ways is often useful in finding efficient algorithms and
error-checking. In this regard we believe spherical field calculations
should be useful in lattice field calculations and vice versa.

\TeXButton{vspace}{\vspace{12pt}}

\noindent \TeXButton{TeX field}
{\vspace{12pt}
\Large\bf{Acknowledgements}
\normalsize\rm}

The author wishes to thank George Brandenburg for use of computing
facilities, Jong Lee for discussions on numerical methods, Sidney Coleman
and Arthur Jaffe for discussions on $\phi ^4$ theory in two dimensions, and
Howard Georgi for numerous conversations and guidance through all stages of
this project.

\newpage

\noindent
Figure 1. The two-loop self-energy diagram for $\phi _0$.
\\
\noindent
Figure 2. Plot of the two-loop self-energy, with normalization given in (\ref
{hh}), for $J_{\max }=0,1,2,3,\infty $.
\\
\noindent
Figure 3. The only divergent diagram, which can be removed by normal
ordering.
\\
\noindent
Figure 4. Plots for $\Delta (t)$ as calculated by finite difference (fd) and
Monte Carlo (mc) methods, with $J_{\max }=0,1$ and $c=0.20.$
\\
\noindent
Figure 5. Plots for $-\Pi (k^2)$ with $J_{\max }=2$ and $c=0.05,0.10,\cdots
0.40.$
\\
\newpage
\begin{figure}
\epsfbox{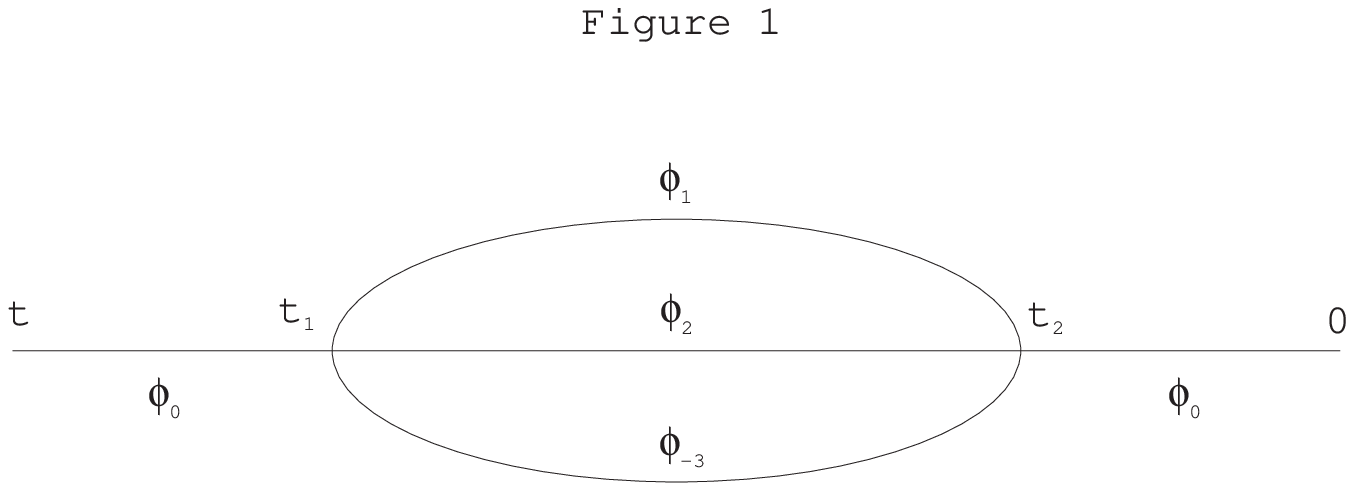}
\end{figure}
\begin{figure}
\epsfbox{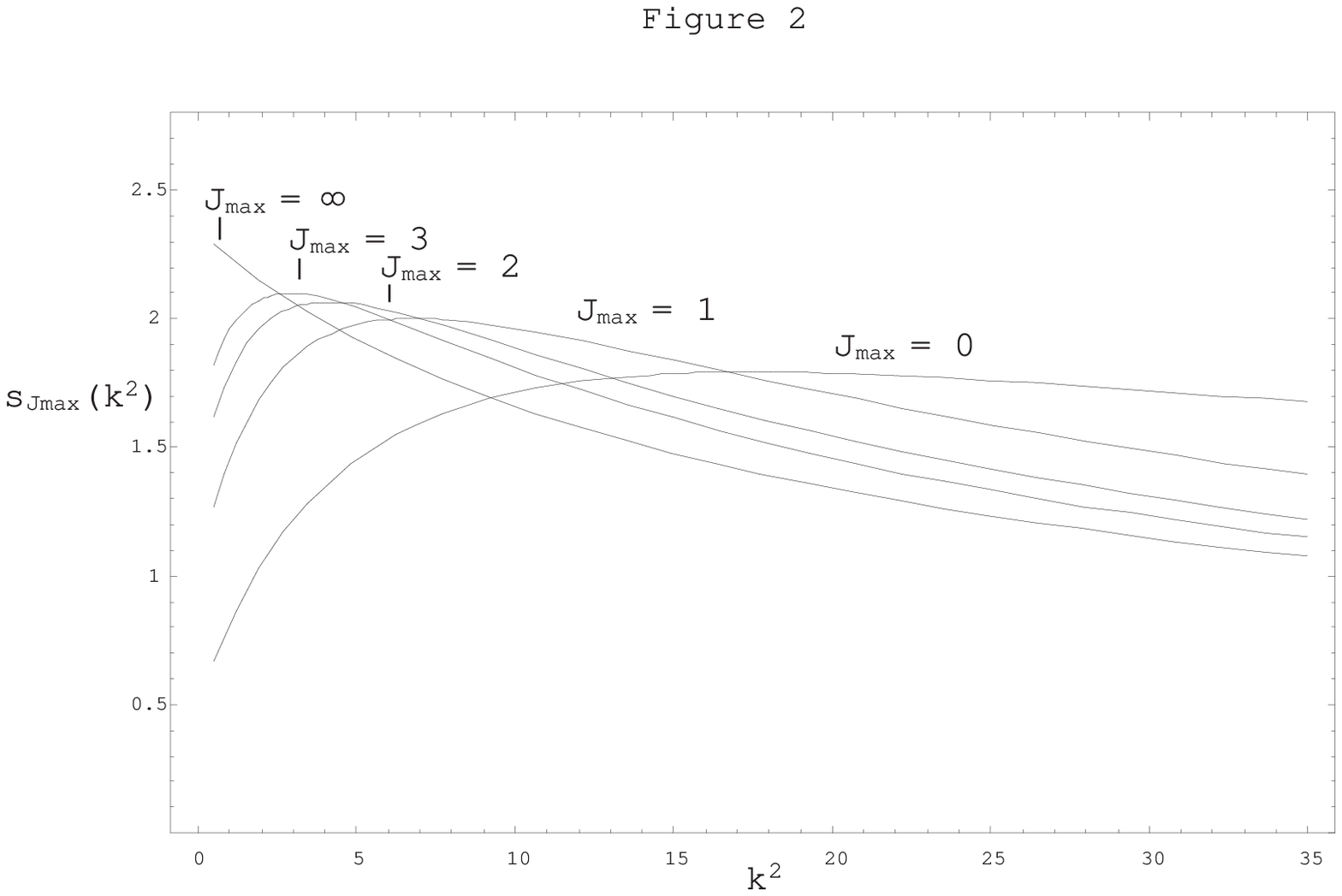}
\end{figure}
\begin{figure}
\epsfbox{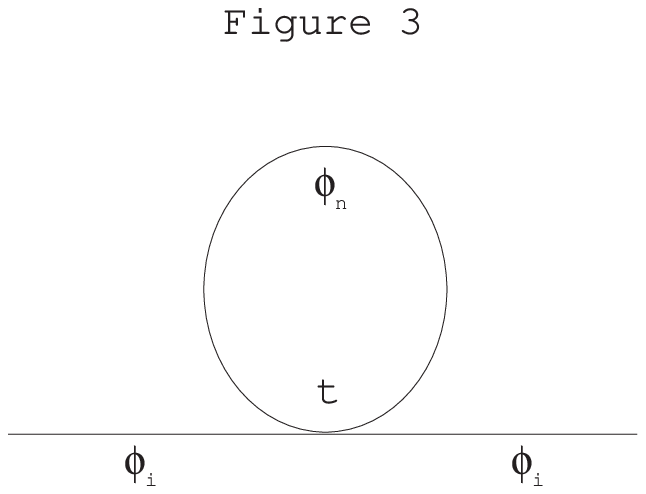}
\end{figure}
\begin{figure}
\epsfbox{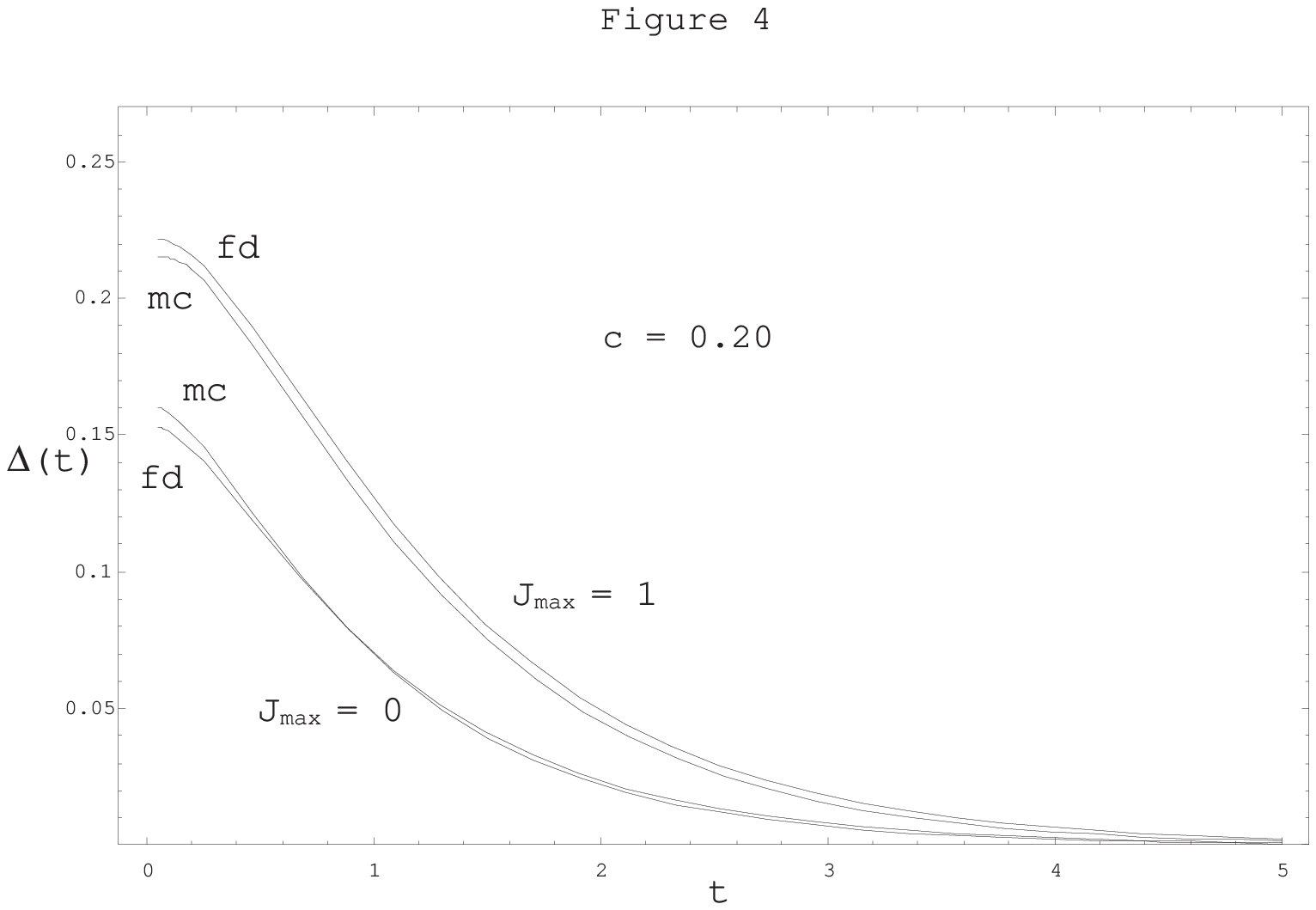}
\end{figure}
\begin{figure}
\epsfbox{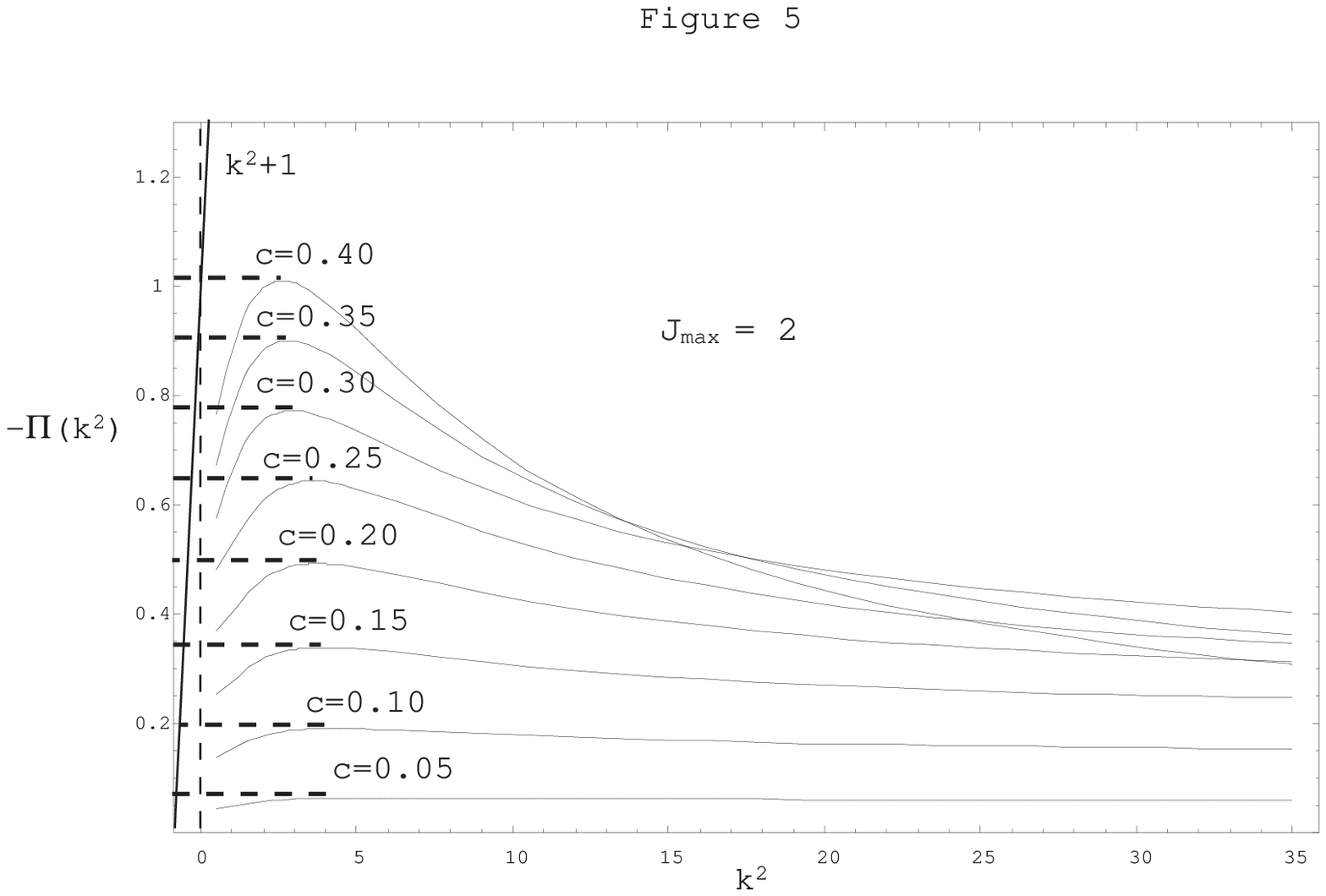}
\end{figure}

\end{document}